\documentclass[showpacs,epsfig,nofootinbib]{article}
\usepackage[margin=1in]{geometry}
\usepackage{adjustbox}
\usepackage{lipsum}
\usepackage{graphicx}
\usepackage{amsfonts}
\usepackage{amsmath}
\usepackage{color}

\usepackage{epstopdf}
\usepackage{latexsym}
\usepackage{amssymb}
\usepackage{amssymb}

\usepackage[center]{subfigure}

\begin{document}

\title{Hilbert repulsion in the Reissner-Nordstr\"{o}m and Schwarzschild spacetimes}

\author{M.-N. C\'{e}l\'{e}rier$^{1}$\footnote{marie-noelle.celerier@obspm.fr}, N. O. Santos$^{1,2,3}$ \footnote{n.o.santos@qmul.ac.uk} \, and V. H. Satheeshkumar$^{3,4}$\footnote{vhsatheeshkumar@gmail.com}\\
\hfill\\
{\small $^{1}$LERMA, CNRS UMR8112, Universit\'e Paris-Sorbonne, Universit\'e Pierre et Marie Curie (UPMC Paris 6),}\\ 
{\small Universit\'e de Cergy-Pontoise, Observatoire de Paris-Meudon 5,} \\ 
{\small Place Jules Janssen, F-92195 Meudon Cedex, France.}\\
\hfill\\
{\small $^{2}$School of Mathematical Sciences, Queen Mary, University of London, London E1 4NS, UK.}\\
\hfill\\
{\small $^{3}$Departamento de F\'{\i}sica Te\'orica, Instituto de F\'{\i}sica,
Universidade do Estado do Rio de Janeiro,}\\
{\small Rio de Janeiro, RJ 20550-900, Brazil.}\\
\hfill\\
{\small $^{4}$Departamento de Astronomia, Observat\'{o}rio Nacional, Rio de Janeiro, RJ 20921-400, Brazil.}
}

 \newcommand{\bq}{\begin{equation}}
 \newcommand{\eq}{\end{equation}}
 \newcommand{\bqn}{\begin{eqnarray}}
 \newcommand{\eqn}{\end{eqnarray}}
 \newcommand{\nb}{\nonumber}
 \newcommand{\lb}{\label}



\maketitle

\abstract{
Studying particle motion in the gravitational field of a black hole from the perspective of different observers is important for separating the coordinate artifacts from the physical phenomena.
In this paper, we show that a freely falling test particle exhibits gravitational repulsion by a black hole as seen by an asymptotic observer, whereas nothing of the kind happens as recorded by a freely falling observer or by an observer located at a finite distance from the event horizon. This analysis is carried out for a general Reissner-Nordstr\"{o}m, an extremal Reissner-Nordstr\"{o}m, and a Schwarzschild black hole. We are lead to conclude that the origin of these bizarre results lies in the fact that the quantities measured by the different observers are neither Lorentz scalars nor gauge invariant.
}


\section{Introduction}

The particle trajectories in the gravitational field of a black hole illustrate some essential features of the black hole spacetime. There is an exhaustive literature on this subject, see \cite{Chandrasekhar:1985kt} for an early example. Yet, it is still an active area of research, for example, recently the circular orbits of neutral \cite{Pugliese:2010ps}  and charged particles \cite{Pugliese:2011py}  was investigated in the Reissner-Nordstr\"{o}m field.

Here, we investigate the radial motion of a freely falling particle in the gravitational field of a Reissner-Nordstr\"{o}m black hole from the standpoint of three different observers. The \textit{far-away observer} is an asymptotic observer who resides in a Minkwoski spacetime far away from any gravitational source. The \textit{finite-distance observer} is at a finite distance from the black hole and hence lives in a locally flat spacetime. Perhaps, the latter is the most neglected among the three observers. The \textit{freely-falling observer} is at rest in the local frame throughout the course of motion, even when passing through the horizon, as long as the tidal forces do not kick in~\cite{Moore:2013sra}. This observer has a finite spacetime span because of a temporary gravitational shielding.

By ``repulsion'' we mean the deceleration of a freely-falling massive particle in the gravitational field of a black hole. This was first noticed by David Hilbert \cite{Hilbert}, hence the name ``Hilbert repulsion''. For a more pedagogical introduction on this topic, we refer the reader to {\cite{CSS}}, and, for an historical view, to {\cite{Spallicci}}.

\section{Falling into a Reissner-Nordstr\"{o}m black hole}

The Reissner-Nordstr\"{o}m metric in the Schwarzschild coordinates\footnote{These are time and distances as measured by a far-way observer} is given by
\bqn
ds^2 &=& \left( 1-\frac{2\mu}{r} + \frac{q^2}{r^2}\right) c^2 dt^2 - \left( 1-\frac{2\mu}{r} + \frac{q^2}{r^2}\right)^{-1} dr^2 \nb\\
&& ~~~~~ ~~~~~ ~~~~~ ~~~~~ ~~~~~ - r^2 (d\theta^2 + sin\theta^2 d\phi^2)
\eqn
with $$ \mu = \frac{G M}{c^2},\,\,\,\, q^2 = \frac{G Q^2}{4 \pi \epsilon_0 c^4},$$
where $M$ is the mass and $Q$ is the electric charge of the black hole, and the constants $G$, $c$ and $\epsilon_0$ have their usual meanings. The horizons are located at $r_\pm = \mu \pm \sqrt{\mu^2 - q^2}$, where the external $r_+$ is an event horizon and the internal $r_-$ is a Cauchy horizon. Here we are only concerned with the event horizon.

For simplicity and without loss of generality, we consider the radial motion of a massive particle in the equatorial plane of a Reissner-Nordstr\"{o}m black hole. Then the timelike geodesic  motion of a test particle starting from rest at infinity obeys the following,
\bqn
\label{2}
\frac{dt}{d\tau}  &=& \left( 1-\frac{2\mu}{r} + \frac{q^2}{r^2}\right)^{-1}, \\
\label{3}
\frac{dr}{d\tau}  &=& \mp \left( \frac{2\mu}{r} - \frac{q^2}{r^2}\right)^{\frac{1}{2}}.
\eqn
Here the minus-or-plus sign in the second equation stands for infalling or outgoing test particles respectively. So now we can find the velocity of the freely falling particle as measured by all the observers. The far-away observer would measure the velocity to be
\bqn
\label{4}
\frac{d r}{dt} = - \left( \frac{2 \mu}{r} - \frac{q^2}{r^2} \right)^{\frac{1}{2}} \left( 1-\frac{2\mu}{r} + \frac{q^2}{r^2}\right) ,
\eqn
which is zero on the horizon.  The finite-distance observer measures the proper time interval as $ d t' = \left( 1-\frac{2\mu}{r} + \frac{q^2}{r^2}\right)^{\frac{1}{2}} \,dt,$ and the proper distance as $ d r' = \left( 1-\frac{2\mu}{r} + \frac{q^2}{r^2}\right)^{-\frac{1}{2}} \,dr.$ Hence the velocity of the freely falling particle as recorded by the finite-distance observer is
\bqn
\frac{d r'}{dt'} = - \left( \frac{2 \mu}{r} - \frac{q^2}{r^2} \right)^{\frac{1}{2}}.
\eqn
After restoring the constants, this is equal to $c$ on the horizon. This means that the finite-distance observer sees that the velocity of the test particle approaches the velocity of light as it nears the horizon, which is quite opposite to that of a far-away observer's measurement. But, it is consoling to find out that the speed of light is still the ultimate speed in General Relativity. Of course, the freely-falling observer is at rest in the float-frame throughout the course of motion, even when passing through the horizon, as long as the tidal forces do not kick in~\cite{Moore:2013sra}.

The radial geodesic equations of motion of a test particle in a Reissner-Nordstr\"{o}m field are given by,
\bqn
\label{6}
\frac{d^2 r}{d\tau^2} &=& - \left( \frac{\mu}{r^2} - \frac{q^2}{r^3} \right) \left( 1 - \frac{2 \mu}{r} +  \frac{q^2}{r^2} \right) \left( \frac{d t}{d \tau} \right)^2 \nb \\
&& + \left( \frac{\mu}{r^2} - \frac{q^2}{r^3} \right) \left( 1 - \frac{2 \mu}{r} +  \frac{q^2}{r^2} \right)^{-1} \left( \frac{d r}{d \tau} \right)^2, \\
\label{7}
\frac{d^2 t}{d\tau^2} &=& -2 \left( \frac{\mu}{r^2} - \frac{q^2}{r^3} \right) \left( 1 - \frac{2 \mu}{r} +  \frac{q^2}{r^2} \right)^{-1} \frac{d t}{d \tau} \frac{d r}{d \tau}.
\eqn
Upon substituting Eq.(\ref{2}) and Eq.(\ref{3}) in Eq. (\ref{6}), we obtain
\bqn
\frac{d^2 r}{d\tau^2} &=& - \frac{1}{r^2} \left( {\mu} - \frac{q^2}{r} \right).
\eqn
Before proceeding further, we would like to point out that, according to the above equation, the motion of a neutral particle is affected by the charge of the black hole, eventually producing a repulsive force for sufficiently small values of $r$. This was noticed earlier in \cite{Barbachoux:2002dq}.

For a freely falling test particle whose proper time is $\tau$, the following relation holds,
\bqn
\frac{dr}{d\tau} = \left( \frac{dr}{dt} \right) \frac{dt}{d\tau}.
\eqn
Further, we have
\bqn
\frac{d^2 r}{d\tau^2} = \frac{d^2 r}{dt^2} \left( \frac{dt}{d\tau} \right)^2 +  \frac{dr}{dt} \, \frac{d^2 t}{d\tau^2},
\eqn
where $\frac{dr}{dt}$ and $\frac{d^2r}{dt^2}$ are the velocity and acceleration of the freely falling particle as recorded by the far-way observer. Substituting for $\frac{d^2 r}{d\tau^2}$ and $\frac{d^2 t}{d\tau^2}$ from the geodesic equations, we get
\bqn
\label{11}
\frac{d^2 r}{dt^2} = \left( \frac{\mu}{r^2} - \frac{q^2}{r^3} \right) \left[ \frac{3}{ \left( 1-\frac{2\mu}{r} + \frac{q^2}{r^2}\right)} \left( \frac{dr}{dt} \right)^2  -  \left( 1-\frac{2\mu}{r} + \frac{q^2}{r^2}\right) \right],
\eqn
From this, we can see that the acceleration of the freely falling particle is positive if
\bqn
\frac{d r}{dt} > \frac{1}{\sqrt{3}} \left( 1-\frac{2\mu}{r} + \frac{q^2}{r^2}\right).
\eqn
In the case of a test particle starting from rest at infinity and observed by an asymptotic observer, Eq.(\ref{11}) can be simplified by substituting Eq.(\ref{4}) to give the following
\bqn
\frac{d^2 r}{dt^2} = - \left( \frac{\mu}{r^2} - \frac{q^2}{r^3} \right) \left( 1-\frac{2\mu}{r} + \frac{q^2}{r^2}\right) \left[ 1 - 3 \left( \frac{2\mu}{r} - \frac{q^2}{r^2}\right) \right]. \nb\\
\eqn
Similarly, the acceleration of a freely falling test particle as observed by a finite-distance observer is given by
\bqn
\frac{d^2 r'}{dt'^2} =- \left( \frac{\mu}{r^2} - \frac{q^2}{r^3}\right) \left( 1 - \frac{2\mu}{r} + \frac{q^2}{r^2}\right)^{\frac{1}{2}}
\eqn
These results are summarized in TABLE I.

\begin{table*}[htp]
\lb{table1}
\caption{Three views of falling into a Reissner-Nordstr\"{o}m black hole.}
\makebox[\textwidth]{
\begin{tabular*}{1.2\textwidth}{l c c c}
\hline
\hline
Quantity &  Far-away Observer & Finite-distance Observer & Freely-falling Observer\\
\hline
Existence & Far-away & Outside Horizon & Everywhere \\
Time & $dt$ & $dt' = \left( 1 - \frac{2\mu}{r} + \frac{q^2}{r^2}\right)^{\frac{1}{2}} dt$ & $d\tau$ \\
Distance & $dr$ & $dr' = \left( 1 - \frac{2\mu}{r} + \frac{q^2}{r^2}\right)^{-\frac{1}{2}} dr$ & $c\,d\tau$ \\
Velocity & $-\left( \frac{2\mu}{r} - \frac{q^2}{r^2}\right)^{\frac{1}{2}} \left( 1 - \frac{2\mu}{r} + \frac{q^2}{r^2}\right)$ & $-\left( \frac{2\mu}{r} - \frac{q^2}{r^2}\right)^{\frac{1}{2}}$ & 0 \\
Velocity at horizon & 0 & $\rightarrow c$ & 0 \\
Maximum Velocity  & $-\left( \frac{2\mu}{r_{*}} - \frac{q^2}{r_{*}^2}\right)^{\frac{1}{2}} \left( 1 - \frac{2\mu}{r_{*}} + \frac{q^2}{r_{*}^2}\right)$ & $-c$ & 0 \\
Acceleration  & $- \left( \frac{\mu}{r^2} - \frac{q^2}{r^3}\right) \left( 1 - \frac{2\mu}{r} + \frac{q^2}{r^2}\right) \Big( 1-  \frac{6\mu}{r} + \frac{3q^2}{r^2} \Big) $ & $- \left( \frac{\mu}{r^2} - \frac{q^2}{r^3}\right) \left( 1 - \frac{2\mu}{r} + \frac{q^2}{r^2}\right)^{\frac{1}{2}} $ & 0 \\
Acceleration at horizon & 0 &  0 &  0 \\
Repulsion  & $r_{*} = 3 \mu + \sqrt{9 \mu^2 - 3 q^2}$ & Never & Never \\
\hline
\hline
\end{tabular*}
}
\end{table*}%

\subsection{General Case}

\begin{figure*}[h]
\includegraphics[width=\textwidth]{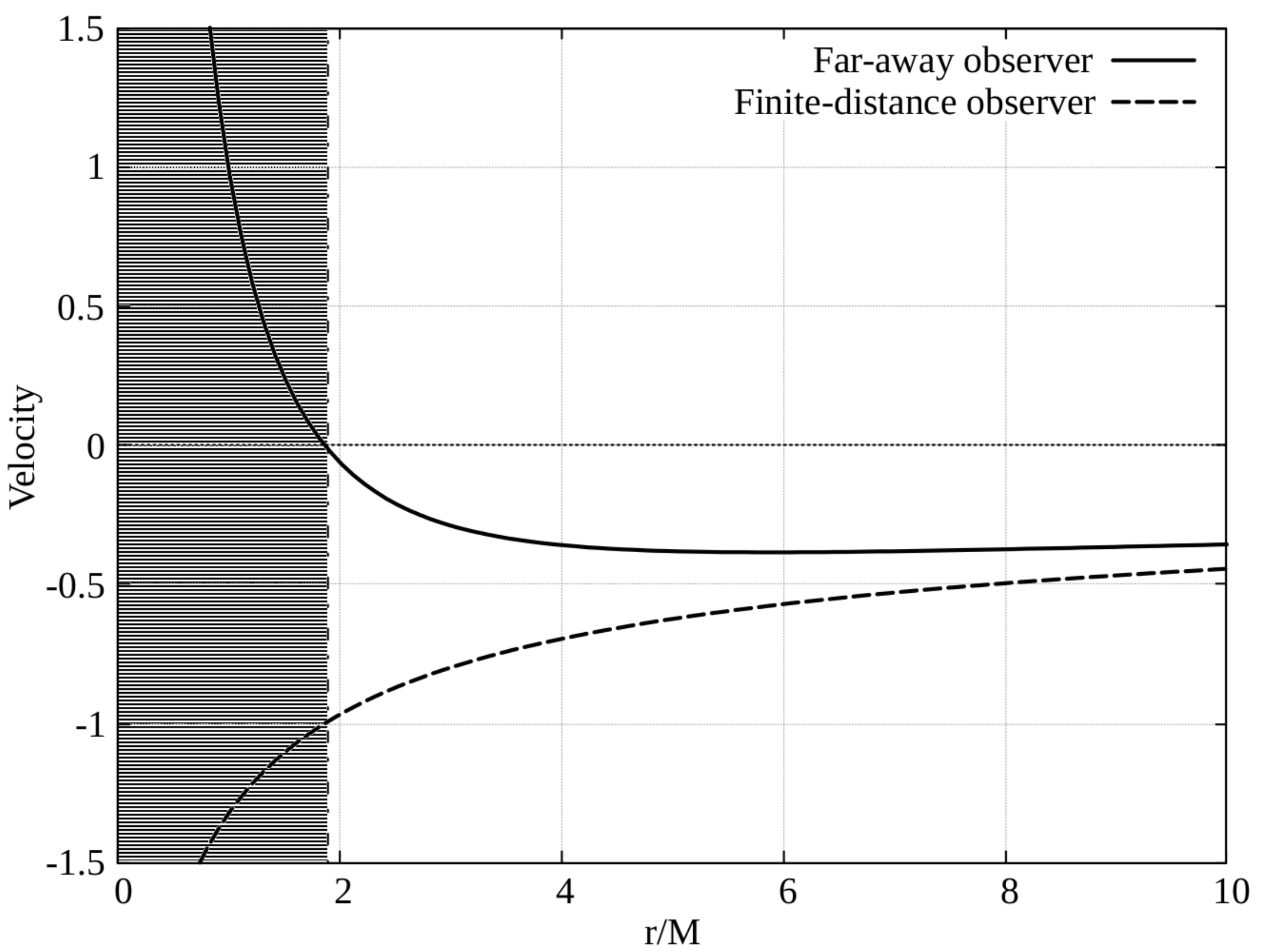}
\caption{Peculiar generic Reissner-Nordstr\"{o}m case ($q=0.5\mu$): the graph shows the variation of the velocity of a freely falling particle as seen by two different observers. Notice that a freely-falling particle crosses the outer horizon ($r_+ = 1.9\mu$) with the speed of light as seen by a finite-distance observer, whereas it stops at this horizon for the far-away observer.}
\label{fig1}
\end{figure*}

\begin{figure*}[h]
\includegraphics[width=\textwidth]{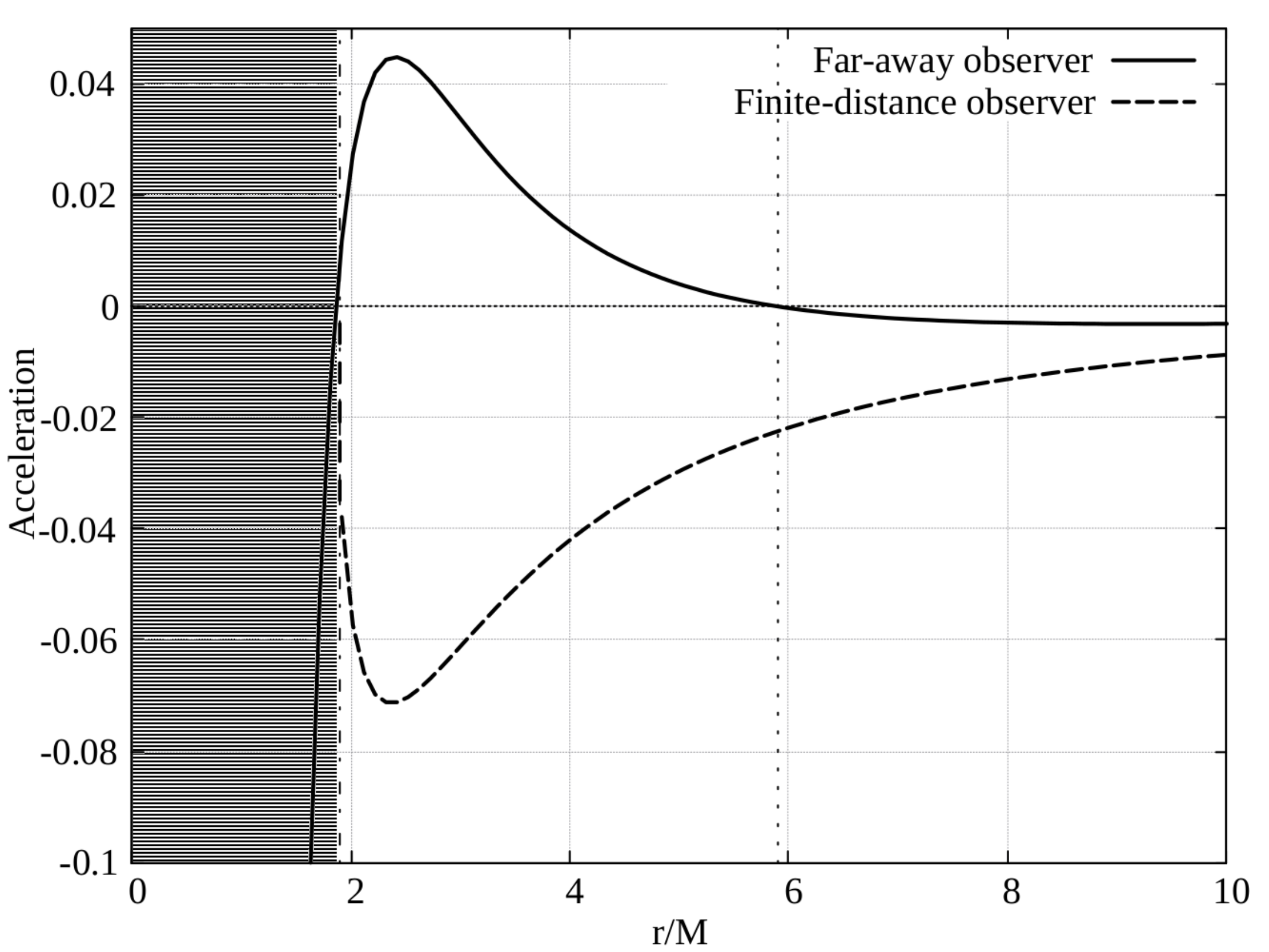}
\caption{Peculiar generic Reissner-Nordstr\"{o}m case ($q=0.5\mu$): the graph shows the variation of the acceleration of a freely falling particle as seen by two different observers. Notice that the acceleration of a freely-falling particle switches sign at $5.9\mu$ as seen by the far-away observer, and is zero at the outer horizon ($r_+ = 1.9\mu$) for both observers.}
\label{fig2}
\end{figure*}

The velocity and acceleration of a particle freely falling into a generic Reissner-Nordstr\"{o}m black hole as  recorded by a far-away observer (fao) are given by
\bqn
v_{fao} &=& - \left( \frac{2 \mu}{r} - \frac{q^2}{r^2} \right)^{\frac{1}{2}} \left( 1-\frac{2\mu}{r} + \frac{q^2}{r^2}\right), \nb\\
a_{fao} &=& - \left( \frac{\mu}{r^2} - \frac{q^2}{r^3} \right) \left( 1-\frac{2\mu}{r} + \frac{q^2}{r^2}\right) \left( 1 -  \frac{6\mu}{r} + \frac{3 q^2}{r^2} \right).\nb\\
\eqn

The only value of $r$ for which $a_{fao}$ switches sign outside the horizon is the greatest zero of the third term in the rhs of the above equation:
\bqn
r_* &=& 3\mu + \sqrt{9\mu^2 - 3 q^2}.
\eqn

The maximum velocity reached at this location is
\bqn
v_{max} &=& - \left[ \frac{2\mu}{3\left(\mu + \sqrt{\mu^2-\frac{q^2}{3}}\right)} - \frac{q^2}{9\left(\mu + \sqrt{\mu^2 - \frac{q^2}{3}}\right)^2}\right]^{1/2} \nb \\
\times && \left[1 - \frac{2\mu}{3\left(\mu + \sqrt{\mu^2-\frac{q^2}{3}}\right)} + \frac{q^2}{9\left(\mu + \sqrt{\mu^2 - \frac{q^2}{3}}\right)^2}\right], \nb\\
\eqn
below which the particle is seen by the far-away observer as decelerating up to the horizon where it stops. Moreover, we have $3\mu \leq r_* < 6\mu$, which is a closer distance than in the Schwarschild case. This means that the presence of a charge is diminishing the effect of the mass of the black hole, which is counterintuitive.

As regards the finite-distance observer (fdo), she records the following:
\bqn
v_{fdo} &=& - \left( \frac{2 \mu}{r} - \frac{q^2}{r^2} \right)^{\frac{1}{2}} \nb \\
a_{fdo} &=& - \left( \frac{\mu}{r^2} - \frac{q^2}{r^3}\right) \left( 1 - \frac{2\mu}{r} + \frac{q^2}{r^2}\right)^{\frac{1}{2}},
\eqn
seeing the particle accelerating all the way to the horizon which it reaches with the speed of light.

Notice also that the acceleration of the free-falling particle vanishes at the outer horizon for both observers.

The variations of the velocity and acceleration of a freely falling particle as seen by each observer are shown in FIG.{\ref{fig1}} and FIG.{\ref{fig2}} respectively, for a given generic Reissner-Nordstr\"{o}m black hole ($q=0.5\mu$).

\subsection{Extremal Case ($q = \mu$)}

\begin{figure*}[h]
\includegraphics[width=\textwidth]{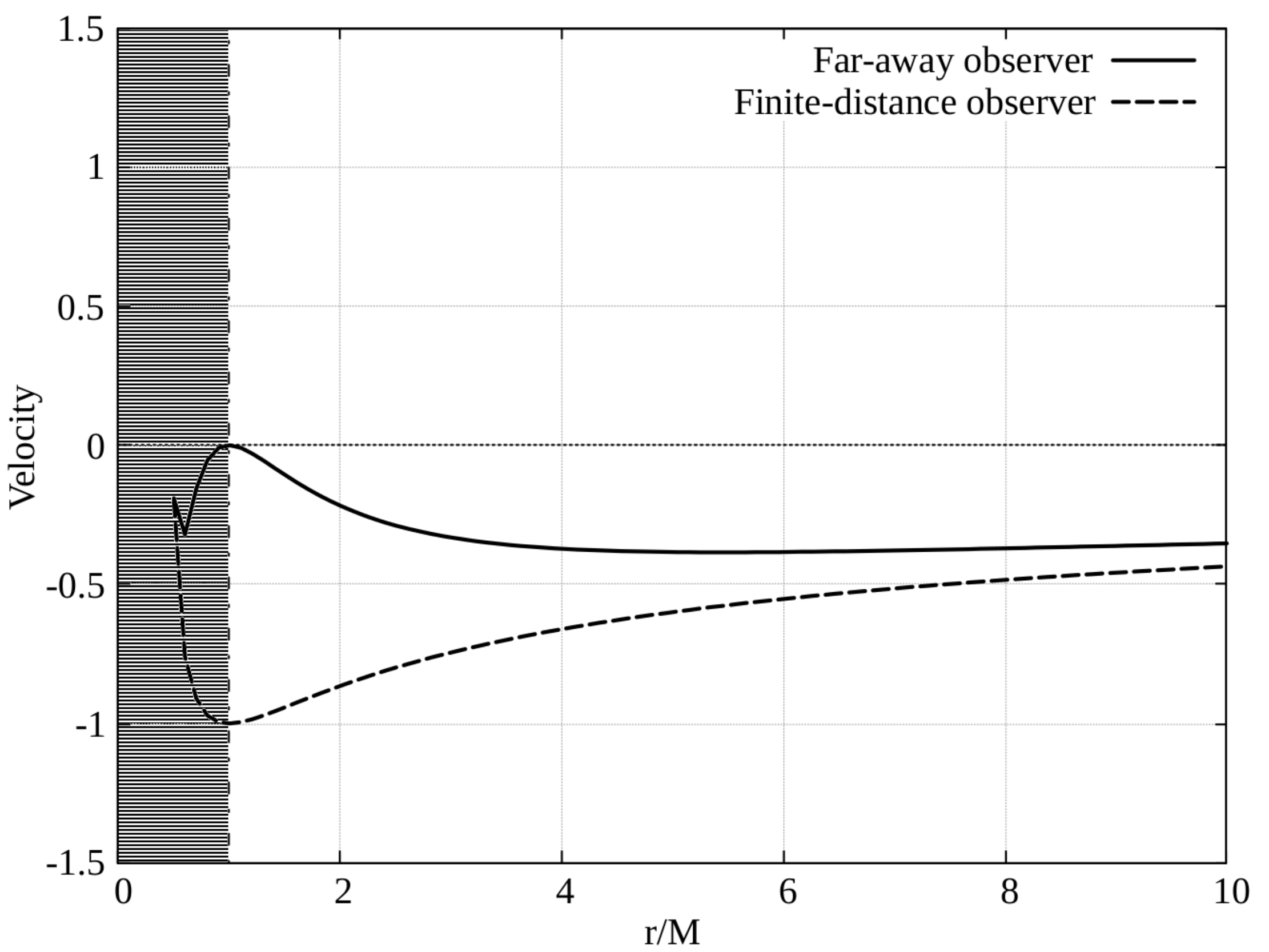}
\caption{Extremal Reissner-Nordstr\"{o}m case ($q=\mu$): the velocities as measured by a far-away and a finite-distance observer. Notice that a freely-falling particle crosses the outer horizon ($r_+ = \mu$) (almost) with the speed of light as seen by the finite-distance observer, whereas it is zero for the far-away observer.}
\label{fig3}
\end{figure*}

\begin{figure*}[h]
\includegraphics[width=\textwidth]{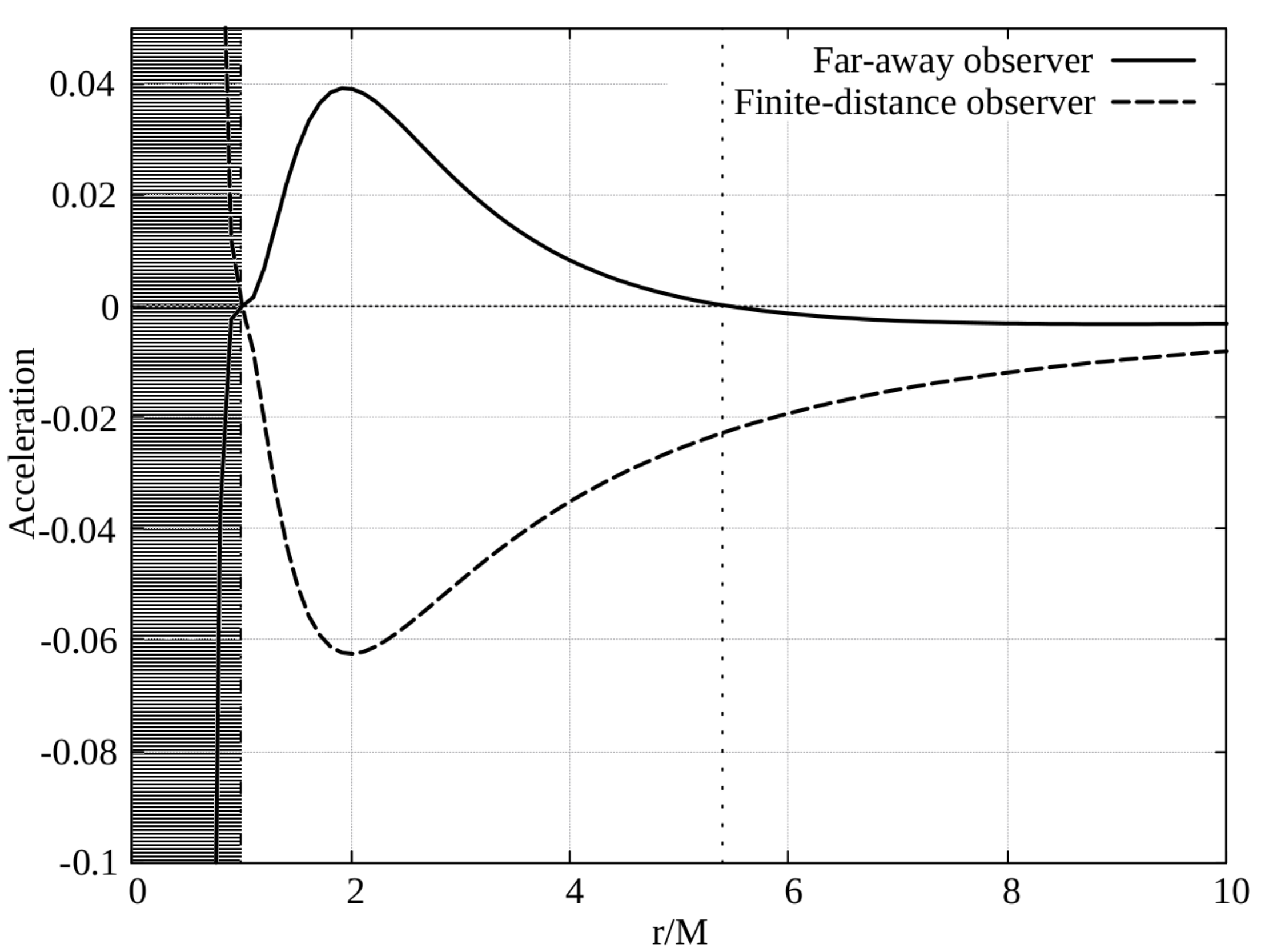}
\caption{Extremal  Reissner-Nordstr\"{o}m case ($q=\mu$): according to a far-away observer, the acceleration of a freely-falling particle switches sign at $5.4\mu$ which is a slightly closer distance than in our peculiar general Reissner-Nordstrom ($5.9\mu$) case and in the Schwarzschild ($6\mu$) case. This means that the presence of a charge is diminishing the effect of the mass of the black hole. This is counterintuitive. The acceleration at the outer horizon ($r_+ = \mu$) is zero for both the observers.}
\label{fig4}
\end{figure*}

The velocity and acceleration of a particle freely falling into an extremal Reissner-Nordstr\"{o}m black hole as  recorded by a far-away observer (fao) are given by
\bqn
v_{fao} &=& - \left( \frac{2 \mu}{r} - \frac{\mu^2}{r^2} \right)^{\frac{1}{2}} \left( 1-\frac{2\mu}{r} + \frac{\mu^2}{r^2}\right), \nb\\
a_{fao} &=& - \left( \frac{\mu}{r^2} - \frac{\mu^2}{r^3} \right) \left( 1-\frac{2\mu}{r} + \frac{\mu^2}{r^2}\right) \left( 1 -  \frac{6\mu}{r} + \frac{3\mu^2}{r^2} \right),\nb\\
\eqn
whereas, a finite-distance observer (fdo) records the following:
\bqn
v_{fdo} &=& - \left( \frac{2 \mu}{r} - \frac{\mu^2}{r^2} \right)^{\frac{1}{2}} \nb \\
a_{fdo} &=& - \left( \frac{\mu}{r^2} - \frac{\mu^2}{r^3}\right) \left( 1 - \frac{2\mu}{r} + \frac{\mu^2}{r^2}\right)^{\frac{1}{2}}.
\eqn

The variations of the velocity and acceleration of a freely falling particle as seen by each observer are shown in FIG.{\ref{fig3}} and FIG.{\ref{fig4}} respectively for an extremal Reissner-Nordstr\"{o}m black hole ($q=\mu$).  Notice that a freely-falling particle crosses the outer horizon ($r_+ = \mu$) (almost) with the speed of light as seen by the finite-distance observer, whereas it stops there for the far-away observer. The acceleration of a freely-falling particle, according to a far-away observer, switches sign at $5.4\mu$ which is still a slightly closer distance than in the Schwarzschild ($6\mu$) case. The acceleration at the outer horizon ($r_+ = \mu$) is zero for both observers.

\subsection{Schwarzschild Case}

\begin{figure*}[h]
\includegraphics[width=\textwidth]{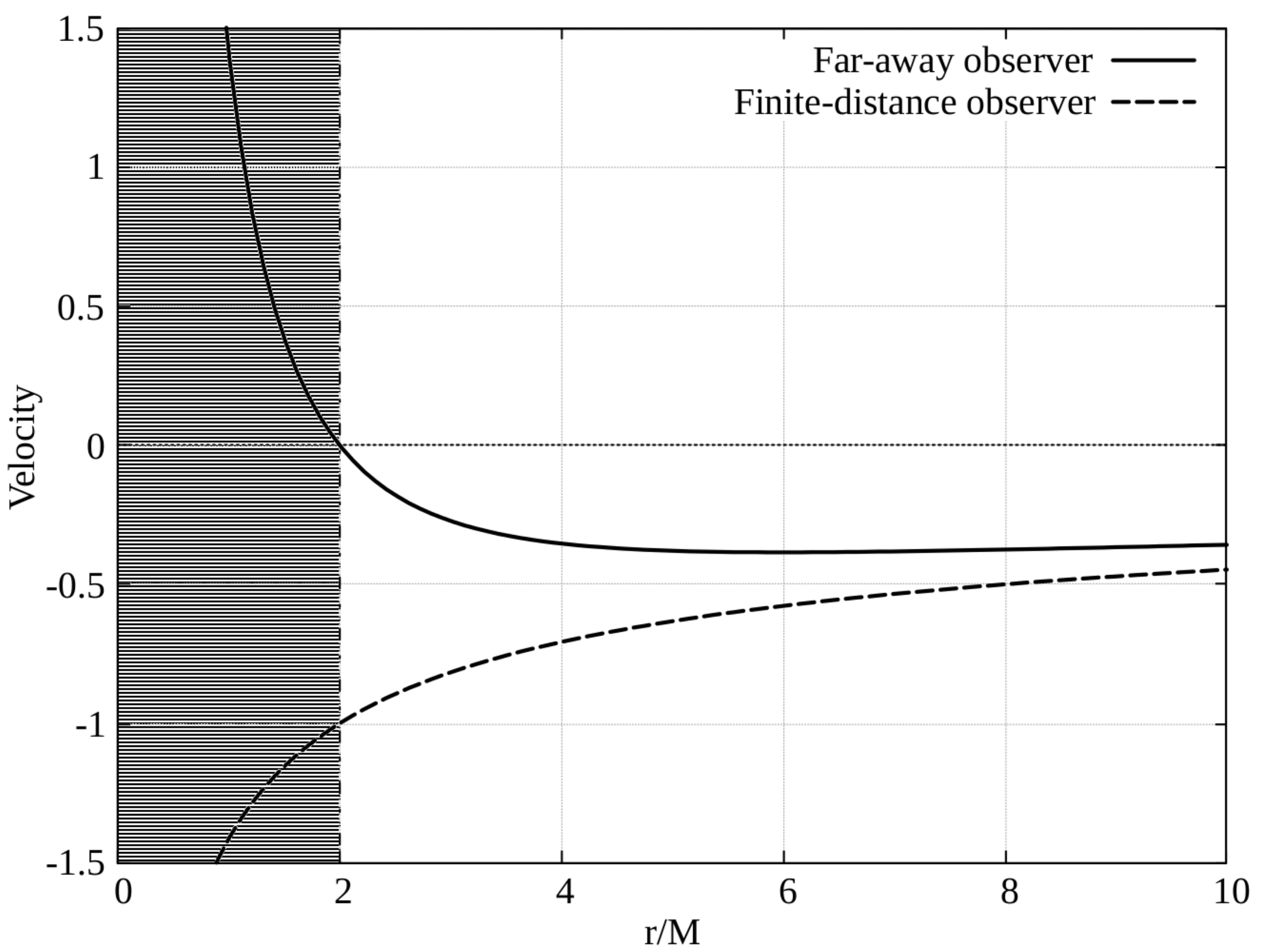}
\caption{Schwarzschild case ($q=0$): the velocities as measured by a far-away and a finite-distance observer. Notice that the velocity of the particle for a far-away observer is zero at the horizon, whereas for a finite-distance observer the freely-falling particle crosses the horizon  with almost the speed of light.}
\label{fig5}
\end{figure*}

\begin{figure*}[h]
\includegraphics[width=\textwidth]{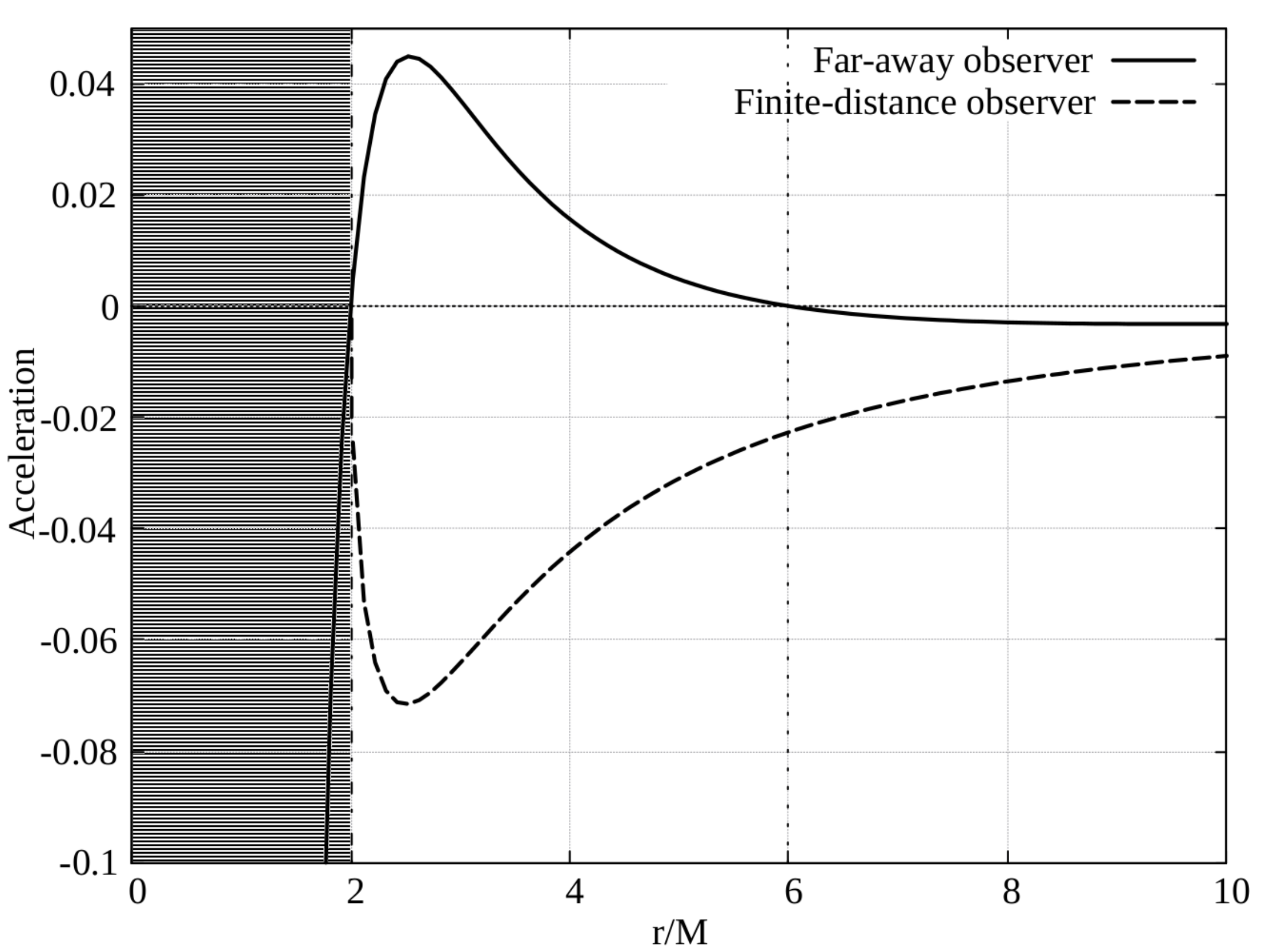}
\caption{Schwarzschild case ($q=0$): the acceleration of a freely-falling particle switches sign (``repulsion") at $r=6\mu$ as measured by a far-away observer, whereas for a finite-distance observer, the acceleration is always negative (attractive). However, the acceleration of the freely-falling particle is zero at the horizon ($r=2\mu$) for both observers.}
\label{fig6}
\end{figure*}

The velocity and acceleration\footnote{This expression was first derived by David Hilbert in 1916 \cite{Hilbert}. This corresponds to Eq. (53) in the original German version, however in the English translation, the corresponding Eq.(63) exhibits a typo in the last term.} of a particle freely falling into a Schwarzschild black hole as  recorded by a far-away observer is given by
\bqn
v_{fao} &=& - \left( \frac{2 \mu}{r} \right)^{\frac{1}{2}} \left( 1-\frac{2\mu}{r} \right), \nb\\
a_{fao} &=& - \left( \frac{\mu}{r^2} \right) \left( 1-\frac{2\mu}{r} \right) \left( 1 -  \frac{6\mu}{r} \right),\nb\\
\eqn
whereas, a finite-distance observer (fdo) records the following:
\bqn
v_{fdo} &=& - \left( \frac{2 \mu}{r} \right)^{\frac{1}{2}} \nb \\
a_{fdo} &=& - \left( \frac{\mu}{r^2} \right) \left( 1 - \frac{2\mu}{r} \right)^{\frac{1}{2}}.
\eqn

The variations of the velocity and acceleration of a freely falling particle as seen by both observers are shown in FIG.{\ref{fig5}} and FIG.{\ref{fig6}} respectively.
The acceleration due to the gravitational field of the black hole is zero at three places, $r=\infty,\, 2\mu \mbox{ (on the horizon)} \,\, \text{and}\,\, 6\mu$. The first two are clear, but it is interesting to see what happens at $r=6\mu$. Acceleration switches sign at that point, and the particle's velocity is maximum and equal to $-2c/3\sqrt{3}$. In summary, as seen by the far-away observer, the freely falling particle accelerates from infinity until it reaches $r=6\mu$, where its acceleration vanishes. From then on, the particle decelerates as it approaches the event horizon. This deceleration in the region $2\mu < r < 6\mu$ is what we call ``repulsion'' by the gravitational field of a black hole.  This is in agreement with the original result of Hilbert~\cite{Hilbert} which was also reproduced much later on by McGruder~\cite{McGruder}.

\section{Conclusion}

Studying the motion of a test particle in the gravitational field of a black hole from the perspective of different observers helps us to distinguish between the artifacts of a particular coordinate system and the real physical effect. We have shown that only a far-away observer sees repulsive gravity i.e., the freely-falling test particle stopping accelerating at a finite distance and starting decelerating to eventually stop at the event horizon. In fact, the far-away observer notices that nothing crosses the event horizon both ways, which makes it a barrier rather than a one-way membrane.  But no such thing is observed by anyone at a finite distance or in free-fall. How finite that distance can be depends on the sensitivity of the observer's instruments \cite{Moore:2013sra}. In summary, an observer at a finite distance sees particles crossing the event horizon and falling into a black hole. In an astrophysical situation, this finite-distance observer can be a space-probe or a satellite detector orbiting a black hole.

One may say that this strange behavior of the gravitational field of a black hole in GR is simply a coordinate effect. But, we know from general coordinate (diffeomorphism) invariance of GR that any observer is as good as any other. What one measures in one's reference frame is as physical as the effect measured in a different frame. But what is important is that all of them are governed by the same laws of physics, i.e., Einstein's field equations. The origin of these bizarre results lies in the fact that the quantities measured by different observers that we are comparing are neither Lorentz scalars nor gauge-invariant.

\section*{Acknowledgments}
We are grateful to Maria de F\'{a}tima Alves da Silva for facilitating this work. VHS was supported by PCI-DB Fellowship from CNPq at the initial stages of this work, and is currently supported by FAPERJ though Programa P\'{o}s-doutorado Nota 10. VHS would like to thank Anzhong Wang and Sofiane Faci for several enlightening discussions, and Jailson Alcaniz for the hospitality at Observat\'{o}rio Nacional.



\begin{thebibliography}{nbound}

\bibitem{Chandrasekhar:1985kt}
  S.~Chandrasekhar,
  \textit{The mathematical theory of black holes,}
(Oxford University Press, Oxford, 1983)

\bibitem{Pugliese:2010ps}
  D.~Pugliese, H.~Quevedo and R.~Ruffini,
  Phys.\ Rev.\ D {\bf 83}, 024021 (2011)
  doi:10.1103/PhysRevD.83.024021
  [arXiv:1012.5411 [astro-ph.HE]].

\bibitem{Pugliese:2011py}
  D.~Pugliese, H.~Quevedo and R.~Ruffini,
  Phys.\ Rev.\ D {\bf 83}, 104052 (2011)
  doi:10.1103/PhysRevD.83.104052
  [arXiv:1103.1807 [gr-qc]].

\bibitem{Moore:2013sra}
  D.~G.~Moore and V.~H.~Satheeshkumar,
  Int.\ J.\ Mod.\ Phys.\ D {\bf 22}, 1342026 (2013)
  [arXiv:1305.7221 [gr-qc]].

\bibitem{Hilbert}
  Hilbert, D.,
  {\em Nachrichten König. Gesell. Wiss. G\"{o}ttingen, Math.-Phys. Kl.}, 53 (1917);
  Hilbert, D.,
  {\em Math. Ann.} 92, 1 (1924).
  For English version see Hilbert, D.,  pp 1017--1038 in {\em The Genesis of General Relativity, Volume 4 -   Gravitation in the Twilight of Classical Physics: The Promise of Mathematics}, edited by J\"{u}rgen Renn and Matthias Schemmel (Springer, Dordrecht. 2007).

\bibitem{CSS} M.~-N.~C\'{e}l\'{e}rier, N.~O.~Santos and V.~H.~Satheeshkumar,
  ``Can gravity be repulsive?''
  [arXiv:1606.08300].

\bibitem{Spallicci}
  A.~Spallicci,
  Fundam.\ Theor.\ Phys.\  {\bf 162}, 561 (2011)
  [arXiv:1005.0611 [physics.hist-ph]].

\bibitem{Barbachoux:2002dq}
  C.~Barbachoux, J.~Gariel, G.~Marcilhacy and N.~O.~Santos,
  Int.\ J.\ Mod.\ Phys.\ D {\bf 11}, 1255 (2002)
  [gr-qc/0203011].

\bibitem{McGruder}
  C.~H.~McGruder,
  Phys.\ Rev.\ D {\bf 25}, 3191 (1982).

\end{thebibliography}
\end{document}